\RequirePackage[2020-02-02]{latexrelease}
\documentclass[%
12pt,
aip,
cha,
amsmath,amssymb,amsthm,
reprint,
floatfix,
]{revtex4-2}

\usepackage[utf8]{inputenc}
\usepackage[T1]{fontenc}
\usepackage{graphicx}
\usepackage{hyperref}
\usepackage{enumitem}
\usepackage{color}
\usepackage{mathtools}

\newtheorem{lemma}{Lemma}

\newcommand{\defn}{\coloneqq} 
\newcommand{\st}{:}  
\newcommand{\isvec}[1]{#1}
\newcommand{\ismat}[1]{\mathbf{#1}}
\newcommand{\isset}[1]{\mathbb{#1}}
\newcommand{\isoperator}[1]{\mathcal{#1}}
\newcommand{\prob}{\operatorname{\mathrm{Pr}}}
\newcommand{\expec}{\operatorname{\mathrm{Ex}}}

\newcommand{\indic}{\mathbf{1}} 

\def\sar{\emph{Sargassum}}
\def\trem{remaining time}

\let\oldleft\left
\let\oldright\right
\renewcommand{\left}{\mathopen{}\mathclose\bgroup\oldleft}
\renewcommand{\right}{\aftergroup\egroup\oldright}

\begin{document}

\title{Stability of temporal statistics in Transition Path Theory with sparse data}

\author{G.\ Bonner}
\email{gbonner@miami.edu}
\affiliation{Department of Atmospheric Sciences, Rosenstiel School of Marine, Atmospheric, and Earth Science, University of Miami, Miami, Florida, USA}

\author{F.J.\ Beron-Vera}
\email{fberon@miami.edu}
\affiliation{Department of Atmospheric Sciences, Rosenstiel School of Marine, Atmospheric, and Earth Science, University of Miami, Miami, Florida, USA}

\author{M.J.\ Olascoaga}
\email{jolascoaga@miami.edu}
\affiliation{Department of Ocean Sciences, Rosenstiel School of Marine, Atmospheric, and Earth Science, University of Miami, Miami, Florida, USA}

\date{\today}

\begin{abstract}
Ulam's method is a popular discretization scheme for stochastic operators that involves the construction of a transition probability matrix controlling a Markov chain on a set of cells covering some domain. We consider an application to satellite-tracked undrogued surface-ocean drifting buoy trajectories obtained from the NOAA Global Drifter Program dataset. Motivated by the motion of \sar{} in the tropical Atlantic, we apply Transition Path Theory (TPT) to drifters originating off the west coast of Africa to the Gulf of Mexico.  We find that the most common case of a regular covering by equal longitude--latitude side cells can lead to a large instability in the computed transition times as a function of the number of cells used. We propose a different covering based on a clustering of the trajectory data which is stable against the number of cells in the covering. We also propose a generalization of the standard transition time statistic of TPT which can be used to construct a partition of the domain of interest into weakly dynamically connected regions.
\end{abstract}

\pacs{02.50.Ga; 47.27.De; 92.10.Fj}

\maketitle

\begin{quotation}
Transition Path Theory (TPT) provides a rigorous statistical characterization of the ensemble of trajectories connecting  directly, i.e., without detours, two disconnected (sets of) states in a Markov chain, a stochastic process that undergoes transitions from one state to another with probability depending on the state attained in the previous step. Markov chains can be constructed using trajectory data via counting of transitions between cells covering the domain spanned by trajectories. With sparse trajectory data, the use of regular cells is observed to result in unstable estimates of the total duration of transition paths. Using Voronoi cells resulting from k-means clustering of the trajectory data, we obtain stable estimates of this TPT statistic, which is generalized to frame the remaining duration of transition paths, a new TPT statistic suitable for investigating connectivity.
\end{quotation}

\section{Introduction}
\label{sec:introduction}

\sar{} is a pelagic seaweed which plays a crucial role in the ecosystem of the Sargasso Sea and surrounding areas of the North Atlantic.\citep{Butler-etal-83}  Large rafts of the seaweed drift through the Caribbean and into the Gulf of Mexico before being circulated into the Sargasso Sea by the Gulf Stream where it is replenished yearly.\citep{Milledge-Harvey-16} These \sar{} clumps provide a habitat to a diverse contingent of invertebrate, fish, and other fauna far away from land. In addition, the Sargasso Sea contributes to approximately 7\% of the global net biological carbon pump due to the abundance \sar{} and the community of organisms it houses.\citep{laffoley2011protection} In 2011, islands in the Caribbean Sea and beaches in South Florida were inundated with abnormally large quantities of \sar.\citep{Gower-etal-13} Since then, waves of \sar{} have been reported regularly in these locations as well as in western Africa and northern Brazil. Although beached \sar{} can add nutrients to coastal soils,\citep{laffoley2011protection} it also creates offensive smells and can result in destruction of some habitats.\citep{vanTussenbroek-etal-17} Large scale cleanup of beaches costs millions annually and can negatively impact tourism in the affected regions. 

The study of the transport of \sar{} across the ocean has attracted recent interest. Satellite-tracked drifter trajectory data from the NOAA Global Drifter Program (GDP) \citep{Lumpkin-Pazos-07} has been used to infer the evolution of the density of \sar{}. Since \sar{} tends to remain on the surface of the ocean, windage effects must be taken into account. In \citet{Beron-etal-22-ADV}, motivated by this consideration, it is demonstrated that the motion of undrogued drifters tracks the actual satellite-inferred density of \sar{} more closely than the drogued counterparts, as the drogue causes the drifter to be less affected by wind. To do this, the North Atlantic is discretized into a large number of small cells which define the states of a Markov chain whose transition probability matrix is constructed based on the initial and final locations of drifter trajectory data on a certain time interval. Provided certain technical conditions are satisfied by this Markov chain, Transition Path Theory (TPT) \citep{vanden2006transition, vanden2010transition} can be applied to identify bottlenecks and fluxes between a source and a target state. In \citet{Beron-etal-22-ADV}, by taking the source to be a single cell off the coast of West Africa, and the target to be the Gulf of Mexico, two prominent paths taken by drifters are revealed. The first is a ``direct'' path along the Great Atlantic \sar{} Belt;\citep{Wang-etal-19} the shape of this path is in agreement with with the satellite-derived density of \sar{} in this region. The second is an ``indirect" southern path whereby drifters circulate toward the Gulf of Guinea before eventually travelling westward along the coast of northern Brazil into the Caribbean.\citep{Franks-etal-16} 
 
The time taken to transition between the source and target is another statistic that can be computed using TPT. It was noticed that the method of \citet{Beron-etal-22-ADV} provides a transition time which is highly sensitive to the number of boxes chosen to cover the domain, creating some distrust in the results. This motivated the development of a new kind of covering which leads to transition times that are stable as a function of the number of cells in the partition. Briefly, this involves clustering the data and generating a covering based on the boundaries of the clusters. Thus, requesting a finer grid tends to result in the division of larger cells while leaving distant cells unchanged. Since we take stability of the transition time as a metric for the trustworthiness of the application of TPT, we review this statistic by proposing a more general one. This new statistic has the properties that it 1) gives the standard transition time as a special case (so long as the source is a single cell) and 2) provides means for partitioning the flow domain to investigate connectivity.

The remainder of this paper is organized as follows. In Section~\ref{sec:background}, we review the theoretical framework of the discretization scheme we apply to the trajectory data in our domain. Section~\ref{sec:transition-time-stability-and-coverings} we apply this discretization scheme to obtain a Markov chain suitable for the application of TPT. We compute transition times and other statistics for two standard kinds of coverings based on regular grids of squares and hexagons to understand their shortcomings. We then propose a different discretization scheme based on the k-means clustering algorithm which is then shown to be significantly more stable. We also investigate the effect of the transition time step through which the trajectory data is temporally ``sliced'' for both regular coverings and our new covering. In Section~\ref{sec:generalized-transition-time} we introduce our generalized transition time and demonstrate how it can be used to obtain a partition of our domain into weakly dynamically connected regions. The proof that our generalized transition time reduces to the standard transition time of TPT in the appropriate limit is in Appendix~\ref{sec:proofs}. Finally, Section~\ref{sec:conclusions} summarizes our results and conclusions.

\section{Background}
\label{sec:background}

\subsection{Trajectory discretization}
\label{subsec:trajectories}

We consider data sets consisting of a series of $J$ disconnected trajectories $\isvec{x}_{1}(t), \isvec{x}_{2}(t), \dotsc, \isvec{x}_{J}(t)$ in $\isset{X}$, where $\isset{X}$ is a subset of the 2-sphere. Each trajectory $\isvec{x}_i(t)$ consists of a number of observations regularly spaced in time by $\Delta t$ units. We suppose that each trajectory is generated by the same underlying nondeterministic dynamical map $\isoperator{L}$ which takes elements of $\isset{X}$ to $\isset{X}$-valued random variables on the appropriate probability space equipped with Lebesgue measure $m$. If $\isoperator{L}$ has a stochastic kernel $K(x, y): \isset{X} \times \isset{X} \to \isset{R}^+$ such that $\isoperator{L}(x) \sim K(x, \cdot)$, where $\int_\isset{X}K(\cdot,y)\,dm(y) = 1$, then we can define the Perron--Frobenius operator, also known as a transfer operator, $\isoperator{P}: L^1(\isset{X}) \to L^1(\isset{X})$ as \citep{Lasota-Mackey-94}
\begin{equation}
    \isoperator{P} f(y) = \int_{\isset{X}} f(x) K(x,y) \, \text{d}m(x).
\end{equation}
The Perron--Frobenius operator describes how an initial distribution is pushed forward by the underlying dynamics. We can study the action of $\isoperator{L}$ numerically by discretizing the Perron--Frobenius operator and using the known trajectory data.
The most widely-used discretization scheme is Ulam's method.\citep{ulam1960collection, li1976finite} Let $\{B_1, \dotsc, B_N\}$ be a partition of $\isset{X}$ into disjoint sets and let $\indic_{\isset{B}}(x)$ be the indicator function on the set $\isset{B}$, which gives 1 when $x\in\isset{B}$ and 0 otherwise. Ulam's method can be interpreted as a Galerkin projection \citep{reddy2019introduction} of $\isoperator{P}$ onto the subspace spanned by $\{\indic_{B_1}, \dotsc, \indic_{B_N}\}$. By choosing basis functions $\{m(B_i)^{-1} \indic_{B_i}(x)\}$, we have that the discretization of $\isoperator{P}$ is an $N$-dimensional linear operator $\ismat{P}$ given by a matrix $(P_{ij})$ such that \citep{Miron-etal-19-Chaos}
\begin{equation} \label{eq:pij-transfer}
    P_{ij} = \frac{m\left(B_i\cap\isoperator{L}^{-1}(B_j)\right)}{m(B_i)}.
\end{equation}
The matrix $\ismat{P}$ is a row-stochastic transition probability matrix which is the discrete analogue of $K(x, y)$. Note that the factor $m(B_i)^{-1}$ in the choice of basis functions is what ensures that $P$ is (row) stochasticized. For computational purposes, we approximate Eq.~\eqref{eq:pij-transfer} in terms of the trajectory data as
\begin{equation} \label{eq:pij-boxes}
    P_{ij} \approx \frac{\sum_{\ell = 1}^{J} \sum_{t} \indic_{B_i}(\isvec{x}_\ell(t)) \indic_{B_j}(\isvec{x}_\ell(t + T))}{\sum_{\ell = 1}^{J} \sum_{t} \indic_{B_i}(\isvec{x}_\ell(t)) },
\end{equation}
where $T$ is some multiple of $\Delta t$. This approach has been used in numerous applications. \citep{froyland2014computational, vansebille2012, junge2009discretization, Miron-etal-17, Olascoaga-etal-18, miron2019lagrangian, miron2021transition, Beron-etal-20-Chaos, beron2022dynamical, Olascoaga-Beron-23, Beron-etal-23-JPO} The transition probability matrix $\ismat{P}$ defines a Markov chain on $N$ states such that the $i$th state is thought of as a delta distribution of mass located at the center of $B_i$. In a practical setting, $T$ must be chosen large enough such that the Markov property holds to suitable precision. In summary, the procedure for the translation of trajectory data into a transition probability matrix involves two main degrees of freedom: 1) a choice of covering of the computational domain by disjoint boxes and 2) a choice of $T$. We return to these issues in Sections~\ref{subsec:regular-coverings} and \ref{subsec:the-time-step-T}, respectively. 

\subsection{Transition Path Theory}
\label{subsec:tpt}

We summarize the key results of Transition Path Theory (TPT) here; details can be found in a series of works.\citep{vanden2006transition, vanden2010transition, metzner2006illustration, helfmann2020extending} We use $\prob (\cdot)$ and $\expec [\cdot]$ to indicate probabilities and expectations, respectively. To begin, we consider a discrete Markov chain $(X_n)_{n \in \isset{Z}}$ on a finite state space $\isset{S}$ with row-stochastic transition probability matrix $\ismat{P}$. It is assumed that the Markov chain is both ergodic (irreducible) and mixing (aperiodic), and homogeneous in time. It follows that there exists a unique stationary distribution $\pi$ which satisfies $\pi \ismat{P} = \pi$. We take $X_0 = \pi$ so that our Markov chain is stationary, that is, we have $X_n = \pi \ismat{P}^n = \pi$ for all $n \in \isset{Z}$. We define the first passage time to a set $\isset{D} \subset \isset{S}$ as
\begin{equation}
    \tau_{\isset{D}}^{+}(n) \defn \inf\{k \geq 0 \st X_{n + k} \in \isset{D} \},
\end{equation}
and the last exit time from $\isset{D}$ as
\begin{equation}
    \tau_{\isset{D}}^{-}(n) \defn \inf\{k \geq 0 \st X_{n - k} \in \isset{D} \}.
\end{equation}
The last exit time is a stopping time with respect to the time-reversed process $(X_{-n})_{n \in Z}$, namely, the Markov chain on $\isset{S}$ with transition probability matrix $\ismat{P}^{-} = (P_{ij}^{-})$ whose entries are given by
\begin{equation}
    P^{-}_{ij} \defn \frac{\pi_j}{\pi_i} P_{ji}.
\end{equation}
Let $\isset{A}, \isset{B}$ be two nonintersecting subsets of $\isset{S}$ such that neither is reachable in one step starting from the other. Following the nomenclature used in physical chemistry literature, at time $n$ we say that the process is forward-reactive $R^{+}(n)$ (respectively, backward-reactive, $R^{-}(n))$ according to the realization of the events 
\begin{equation}
    R^{\pm}(n) \defn \{ \tau_{\isset{B}}^{\pm}(n) < \tau_{\isset{A}}^{\pm}(n) \}.
\end{equation}
Then, the process is reactive at time $n$ if
\begin{equation}
    R(n) \defn \left\{R^{-}(n) \cup R^{+}(n)\right\}.
\end{equation}
In summary, a trajectory is reactive at time $n$ if its most recent visit to $\isset{A} \cup \isset{B}$ was to $\isset{A}$, it is currently outside of $\isset{A} \cup \isset{B}$, and its next visit to $\isset{A} \cup \isset{B}$ will be to $\isset{B}$. One thinks of $\isset{A}$ as a source and $\isset{B}$ as a target for some process. Associated to the forward and backward reactivities are the forward and backward committors $q_{i}^{\pm}(n)$ defined for $i \in \isset{S}$ by
\begin{equation}
    q_{i}^{\pm}(n) \defn \prob(R^{\pm}(n) \mid X_n = i) .
\end{equation}
One can show via a first step analysis that in the case of a homogeneous and stationary Markov chain, the committors are independent of $n$ and satisfy linear matrix equations
\begin{widetext}
\begin{equation}
    q^{+}_i = \begin{cases}
        \sum_{j \in \isset{S}} P_{ij} q_{j}^{+} & i \notin \isset{A} \cup \isset{B},\\ 
        0 & i \in \isset{A},\\ 
        1 & i \in \isset{B},
    \end{cases} \quad  \text{and} \quad   q^{-}_i = \begin{cases}
        \sum_{j \in \isset{S}} P_{ij}^{-} q_{j}^{-} & i \notin \isset{A} \cup \isset{B},\\ 
        1 & i \in \isset{A},\\ 
        0 & i \in \isset{B}.
    \end{cases}
\end{equation}
\end{widetext}
Using the committors, a number of statistics can be computed for reactive trajectories. First, we have the reactive density
\begin{equation} \label{eq:muAB}
    \mu^{\isset{AB}}_{i}(n) \defn \prob(X_n = i,\, R(n)) = q_{i}^{-} \pi_i q_{i}^+, \quad i \notin \isset{A} \cup \isset{B}.
\end{equation}
States with large large reactive densities relative to their neighbors are interpreted as bottlenecks for reactive trajectories. We also define the reactive current
\begin{widetext}
\begin{equation}
    f^{\isset{AB}}_{i j}(n) \defn \prob(X_n = i,\, R^{-}(n), X_{n + 1} = j,\, R^{+}(n + 1)) = q_{i}^{-} \pi_i P_{ij} q_{j}^+, \quad i, j \in \isset{S},
\end{equation}
\end{widetext}
as well as the effective reactive current
\begin{equation} \label{eq:effective-reactive-current}
    f_{i j}^{+} \defn \max\{f^{\isset{AB}}_{i j} - f^{\isset{AB}}_{j i}, 0\}.
\end{equation}
The effective reactive current is a kind of $\isset{B}$-facing gradient of the reactive density; it identifies pairs of states with a large net flow of probability. Finally, and of particular importance here is the transition time $t^{\isset{AB}}$. The original definition by \citet{vanden2006transition} of $t^{\isset{AB}}$ is as the limiting ratio of the time spent during reactive transitions from $\isset{A}$ to $\isset{B}$ to the rate of reactive transitions leaving $\isset{A}$. In  \citet{helfmann2020extending}, the following expression is provided in the discrete case
\begin{equation} \label{eq:tAB-vanE}
    t^{\isset{AB}} \defn \frac{\prob(R(n))}{\prob(R^+(n + 1), X_n \in A)} = \frac{\sum_{i \in \isset{S}} \mu_{i }^{\isset{AB}}}{\sum_{i \in \isset{A}, j \in \isset{S}} f_{i j}^{\isset{AB}}}.
\end{equation}
In Section~\ref{sec:generalized-transition-time} we will introduce a generalization of the transition time which will allow us to express Eq.~\eqref{eq:tAB-vanE} as a straightforward expectation.

\subsection{Open systems and connectivity}
\label{subsec:open-systems}

In many cases, the trajectory data are given on an open domain, and further processing is required to obtain a suitable Markov chain. We follow \citet{miron2021transition} and subsequent works, and introduce a two-way nirvana state to create a closed system. Suppose that all trajectory data are contained inside a domain $\isset{Y} \subset \isset{X}$. We partition $\isset{Y}$ as $\isset{Y} = \isset{Y}^O \cup \omega$ such that $\partial \isset{Y} \subset \omega$ and $|\omega| \ll |\isset{Y}^O|$. We then construct a covering by $N$ boxes of $\isset{Y}^O$ with one additional box appended corresponding to the whole of $\omega$, the nirvana state. Applying Eq.~\eqref{eq:pij-boxes}, we obtain a row-stochastic transition matrix of the form
\begin{equation}
    \ismat{P} = \begin{pmatrix}
        \ismat{P}^{O \to O} & \ismat{P}^{O \to \omega} \\ 
        \ismat{P}^{\omega \to O} & 0 
    \end{pmatrix},
    \label{eq:Pclose}
\end{equation}
where $\ismat{P}^{O \to O}$ is $N \times N$, $\ismat{P}^{O \to \omega}$ is $N \times 1$ and $\ismat{P}^{\omega \to O}$ is $1\times N$. Note that trajectories which begin and end in the nirvana state are ignored. In general, we are only interested in reactive trajectories which do not visit this extra nirvana state. This requirement is equivalent to making the replacements $\isset{A} \to \isset{A} \cup \omega$ and $\isset{B} \to \isset{B} \cup \omega$ in the basic TPT formulae. One can show \citep{miron2021transition} that this is also equivalent to leaving $\isset{A}$ and $\isset{B}$ unchanged, but replacing $\ismat{P}$ with the row-substochastic matrix $\ismat{P}^{O \to O}$ and $\pi$ by restriction of the stationary distribution of Eq.\@~\ref{eq:Pclose} to $O$. We apply the later method due to convenience of computation.

Depending on the shape of the data, $\ismat{P}^{O \to O}$ may not an irreducible, aperiodic matrix. To remedy this, we apply Tarjan's algorithm \citep{tarjan1972depth} to extract the largest strongly connected component of $\ismat{P}^{O \to O}$. Then, $\ismat{P}^{O \to O}$ is modified to remove all other states including contributions from trajectories which pass through the removed states. The final result is that we have an irreducible, aperiodic matrix which avoids the nirvana state and is suitable for use in the formulation of TPT above.

\section{Transition time stability and coverings}
\label{sec:transition-time-stability-and-coverings}

\subsection{Regular coverings}
\label{subsec:regular-coverings}

\begin{figure}[t!]
    \centering
    \includegraphics[width = \linewidth]{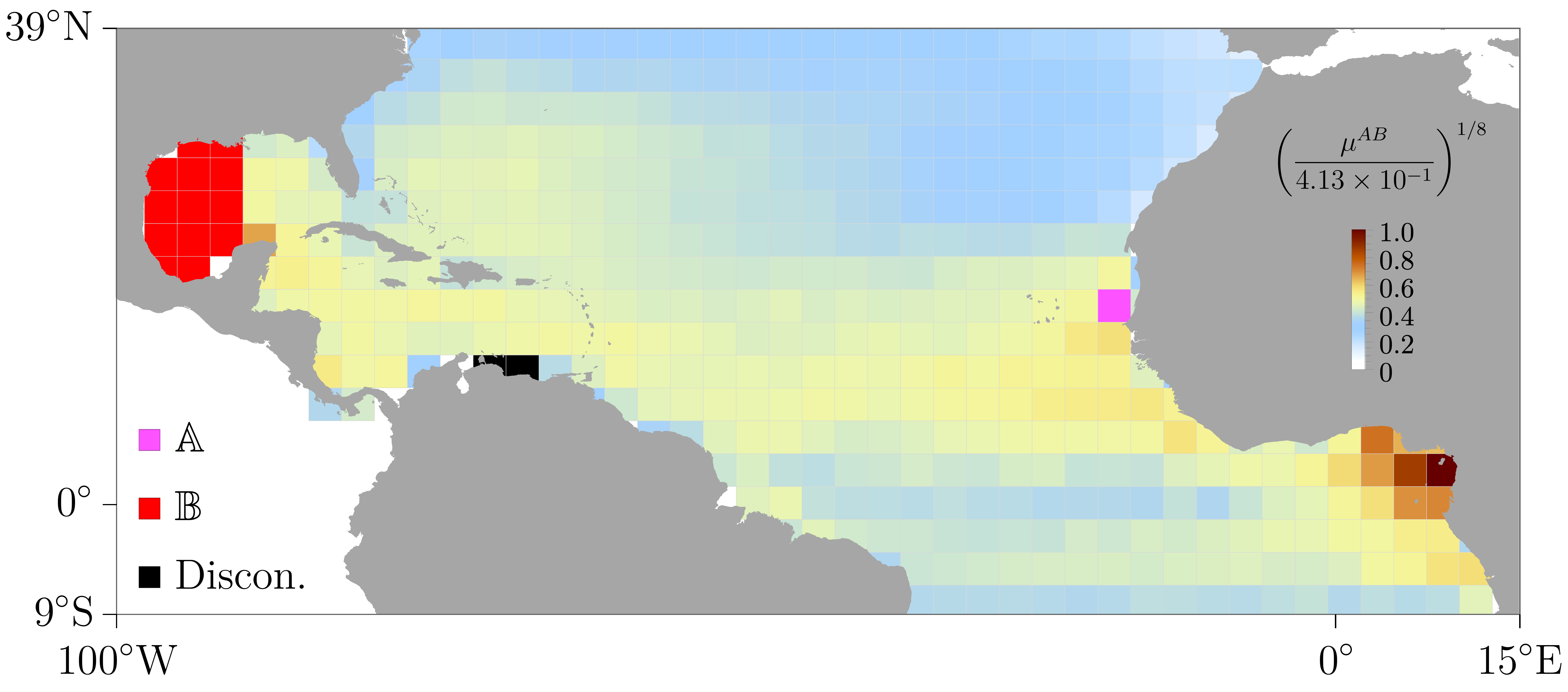}
    \includegraphics[width = \linewidth]{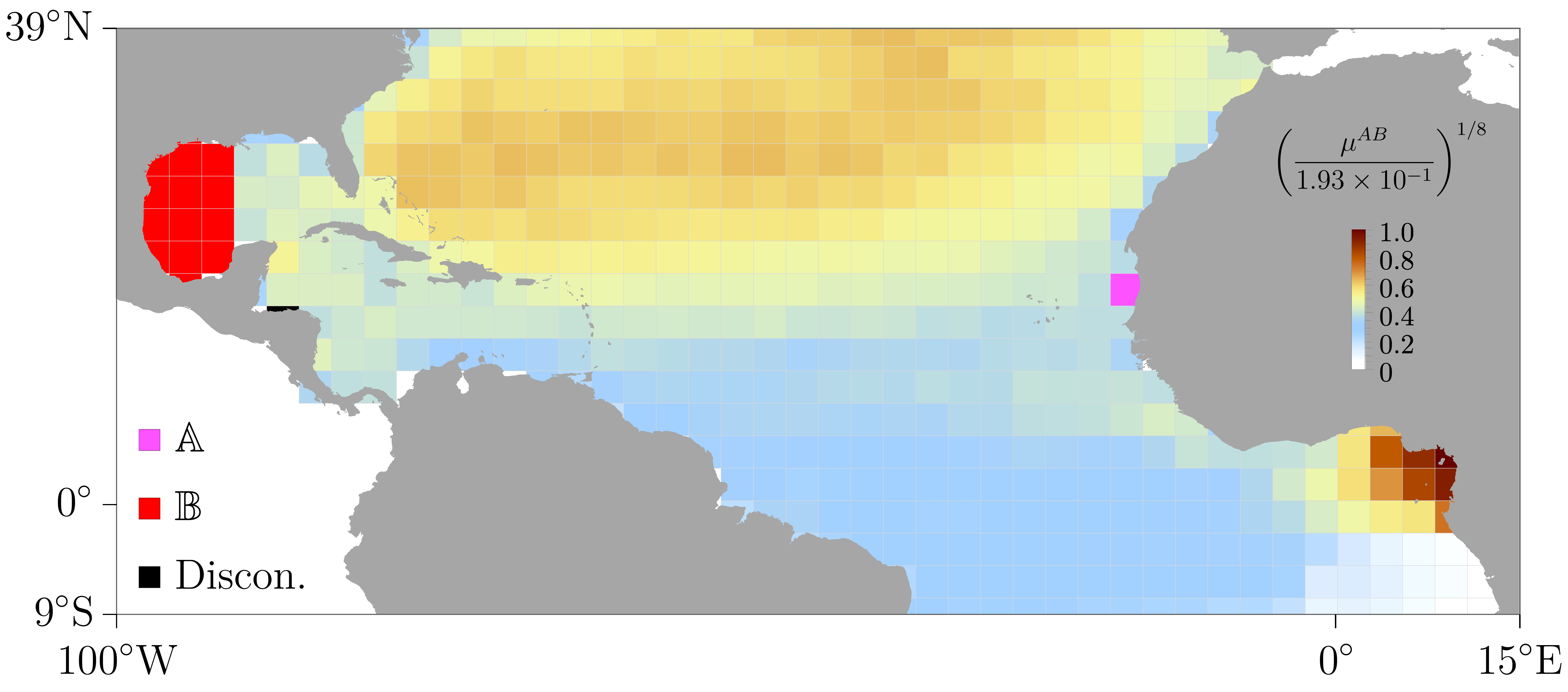}
    \caption{(top panel) The eighth-root transformation of the reactive density $\mu^{\isset{AB}}$ in the North Atlantic constructed from GDP undrogued drifter data. The computational domain was initialized with 760 square boxes. Boxes colored in black contained data but were removed due to being part of a reducible subset of the directed graph associated with the Markov chain resulting from discretizing the drifter motion using Ulam's method. (bottom panel) As in the top panel, but with an initialization of 780 boxes.}
    \label{fig:mu-reg}
\end{figure}

We apply the methods described in Section~\ref{sec:background} to drifter trajectory data obtained from the NOAA Global Drifter Program (GDP).\citep{Lumpkin-Pazos-07} In particular, we use quarter-daily interpolated data of the positions of drifters in the tropical Atlantic. We are primarily interested in undrogued drifter trajectories as these may be more accurate models for the motion of \emph{Sargassum} than their drogued counterparts, as noted in the Introduction. After discarding sections of trajectories which still have their drogue, we must choose a time step $T$, cf.\ Section~\ref{subsec:trajectories}. Following, e.g., \citet{beron2022dynamical}, we choose $T = 5\,\text{days}$, a timescale such longer than the Lagrangian decorrelation time scale for the ocean of $1\,\text{day}$.\citep{lacasce2008statistics} This ensures that the assumption of Markovianity will hold to suitable accuracy. In general, Eq.~\eqref{eq:pij-boxes} is used except where trajectories contain holes or the length of the trajectory is shorter than $T$. After obtaining the transition matrix, we apply Eq.~\eqref{eq:muAB} to calculate the reactive density, choosing $\isset{A}$ concentrated off the coast of West Africa and $\isset{B}$ as the Gulf of Mexico. We cover the computational domain with 760 boxes, resulting in boxes of about $2.4^\circ$ side, of which 463 both contained data and were not disconnected. We will sometimes refer to this partition loosely as a partition into ``squares,'' keeping in mind that the covering actually exists on a 2-sphere. The calculation was then repeated with a covering of 780 boxes. The results are shown in Fig.~\ref{fig:mu-reg}.



In addition to the dramatic difference in the density of $\mu^{\isset{AB}}$, we find that $t^{\isset{AB}} = 9.81\,\text{yr}$ for the coarser partition (Fig.~\ref{fig:mu-reg}, top panel) and $t^{\isset{AB}} = 186\,\text{yr}$ for the finer partition (Fig.~\ref{fig:mu-reg}, bottom panel). In general, changing the number of boxes in the covering results in $\mu^{\isset{AB}}$ graphs which oscillate between patterns similar to the distributions in Fig.~\ref{fig:mu-reg}. We will first address the question of why a small change in the number of boxes in the covering can lead to a large change in these TPT statistics. Note that these large changes are not caused by our choice of $\isset{A}$ or $\isset{B}$, it is an issue caused by the nature of coverings by a regular grid of squares. A fundamental issue with a covering by squares is that an addition of even a small number of boxes can result in a shift in location of every box in the covering. When dealing with sparse trajectory data, this can radically change the outflow in certain regions of the space. This is observed in Figs.~\ref{fig:mu-reg} in the region $[75^\circ\text{W}, 20^\circ\text{W}] \times [20^\circ\text{N}, 39^\circ\text{N}]$. There, the bottom panel of Fig.~\ref{fig:mu-reg} shows a much larger portion of reactive density, apparently suggesting that particles tend to circulate in this area before eventually finding the more direct path from $\isset{A}$ to $\isset{B}$ highlighted in the top panel of Fig.~\ref{fig:mu-reg}.

\begin{figure}[t!]
    \centering
    \includegraphics[width = \linewidth]{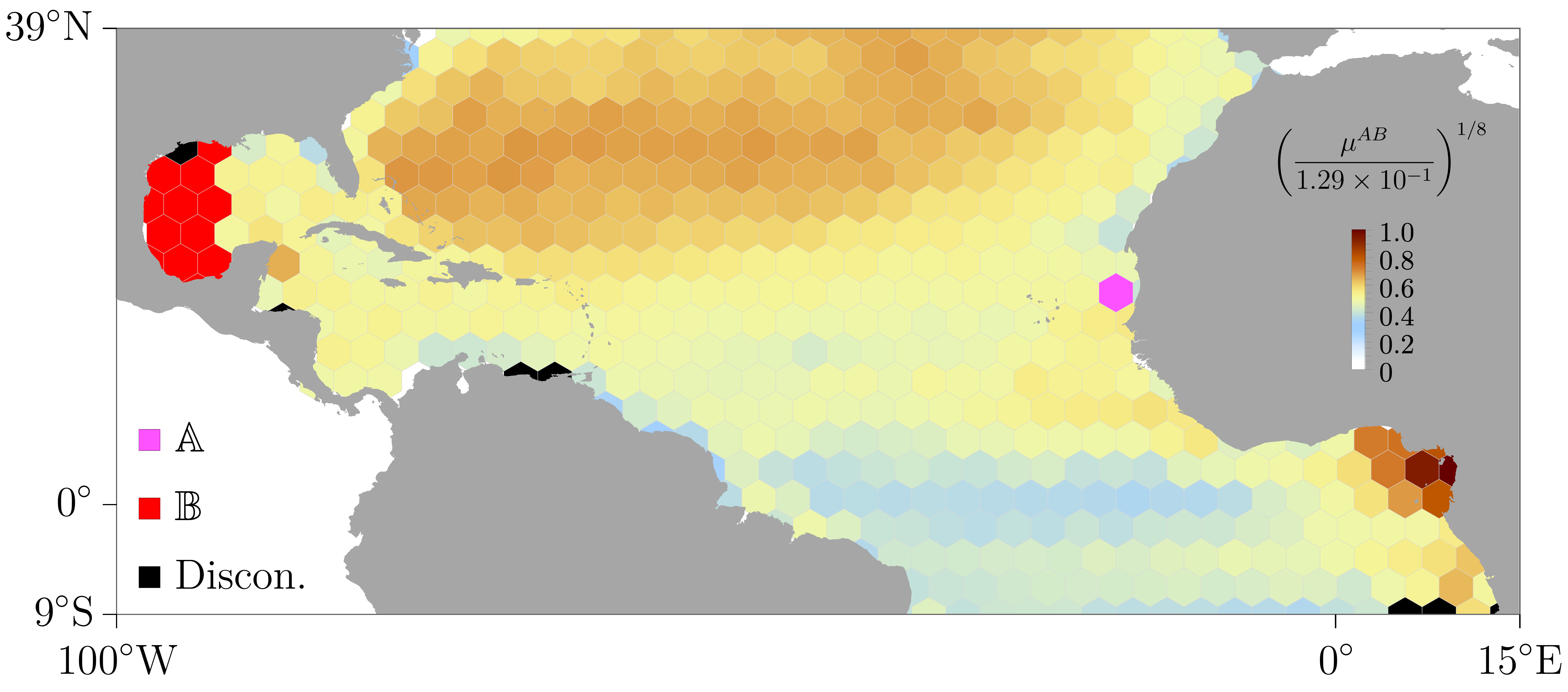}
    \includegraphics[width = \linewidth]{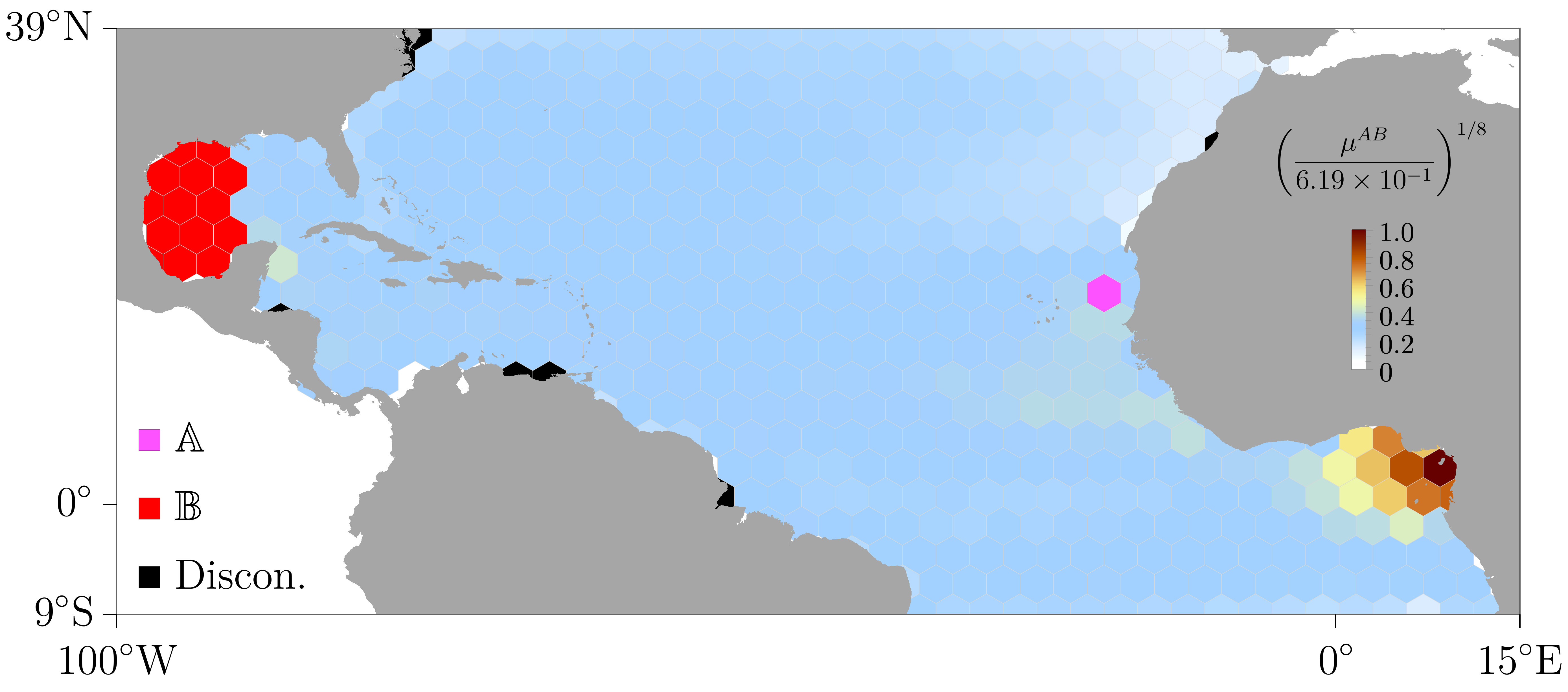}
    \caption{As in Fig.~\ref{fig:mu-reg} but with a hexagonal covering. The computational domain in the top (respectively, bottom) panel was initialized with 820 (respectively, 840) boxes.}
    \label{fig:mu-hex}
\end{figure}

Another option for a regular covering is by hexagons. Hexagons could be considered a more natural choice than squares since the distance between the centers of adjacent hexagons is constant. In \citet{o2021estimating}, a hexagonal covering provided by the H3 spatial index \citep{uberh3} was used to construct the transition matrix. We choose the same parameters as for the squares, but instead cover the computational domain by a regular grid of hexagons. This is repeated twice with a small difference in the number of initial cells; the results are shown in Fig.~\ref{fig:mu-hex}. Again we find that a small change in the number of covering cells leads to a large change in transition path theory statistics. In addition to the differences in the reactive densities, we find $t^{\isset{AB}} = 25.6\,\text{yr}$ for the coarser partition (Fig.~\ref{fig:mu-hex}, top panel) and $t^{\isset{AB}} = 156\,\text{yr}$ for the finer partition (Fig.~\ref{fig:mu-hex}, bottom panel).
The essential problem is the same for both square and hexagonal coverings, namely, that it is not clear which resulting statistics should be trusted. One would hope that variations in a scalar statistic such as $t^{\isset{AB}}$ would settle as the number of boxes increases, but this is not the case with regular coverings. Motivated by these examples, we propose another kind of covering which addresses this issue.

\subsection{Voronoi coverings}
\label{subsec:voronoi-coverings}

\begin{figure}[t!]
    \centering
    \includegraphics[width = \linewidth]{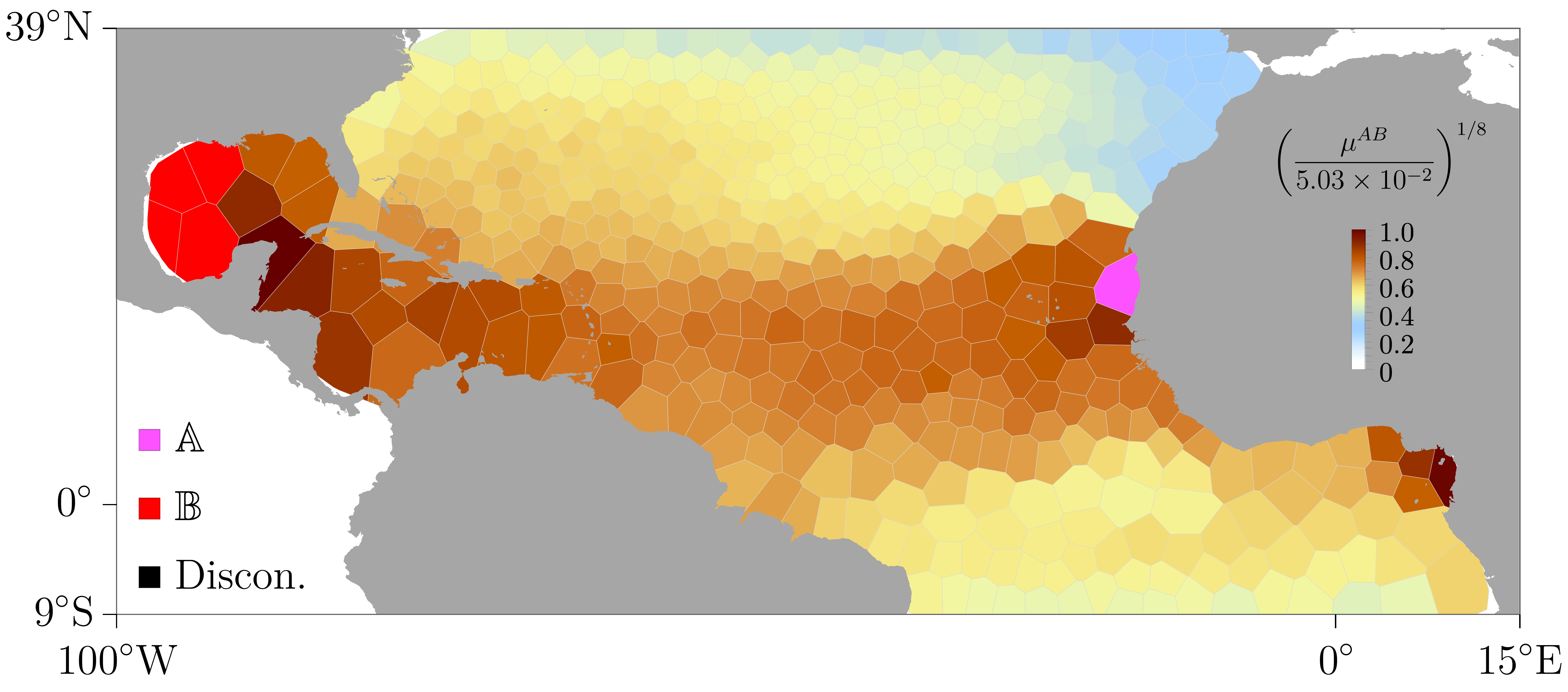}
    \caption{As in Fig.~\ref{fig:mu-reg}, but with a Voronoi covering generated by k-means with 500 clusters.}
    \label{fig:mu-vor-500}
\end{figure}

We propose that instead of covering the domain with a regular grid, we instead cluster the observations and draw polygons based on the cluster boundaries. The intended result should be that data points which are close to each other should tend to end up in the same box and hence the derived transition matrix should be robust against small changes in box number. There are a number of clustering algorithms, for the current application we have chosen to use the k-means method.\citep{forgy1965cluster, lloyd1982least} This is a hard clustering algorithm which is guaranteed to converge and create $n$ clusters (if possible) when requested. In addition, k-means is straightforward to implement and is built into the clustering packages of many popular languages. The output of k-means is a collection of centroids such that the each data point belongs to the cluster defined by its closest centroid in terms of Euclidean distance. Hence, we can define the polygons covering the computational domain using a Voronoi tessellation; cf., e.g., \citet{burrough2015principles}. We compute the intersection of the convex hull of the data with the Voronoi tessellation to reduce the size of the outer cells for clearer visualization. Performing the same reactive density calculation as in Section~\ref{subsec:regular-coverings} gives the results shown in Figure.~\ref{fig:mu-vor-500}. The picture is rather insensitive to the number of boxes considered, which we quantify below.

\begin{figure}[t!]
    \centering
    \includegraphics[width = \linewidth]{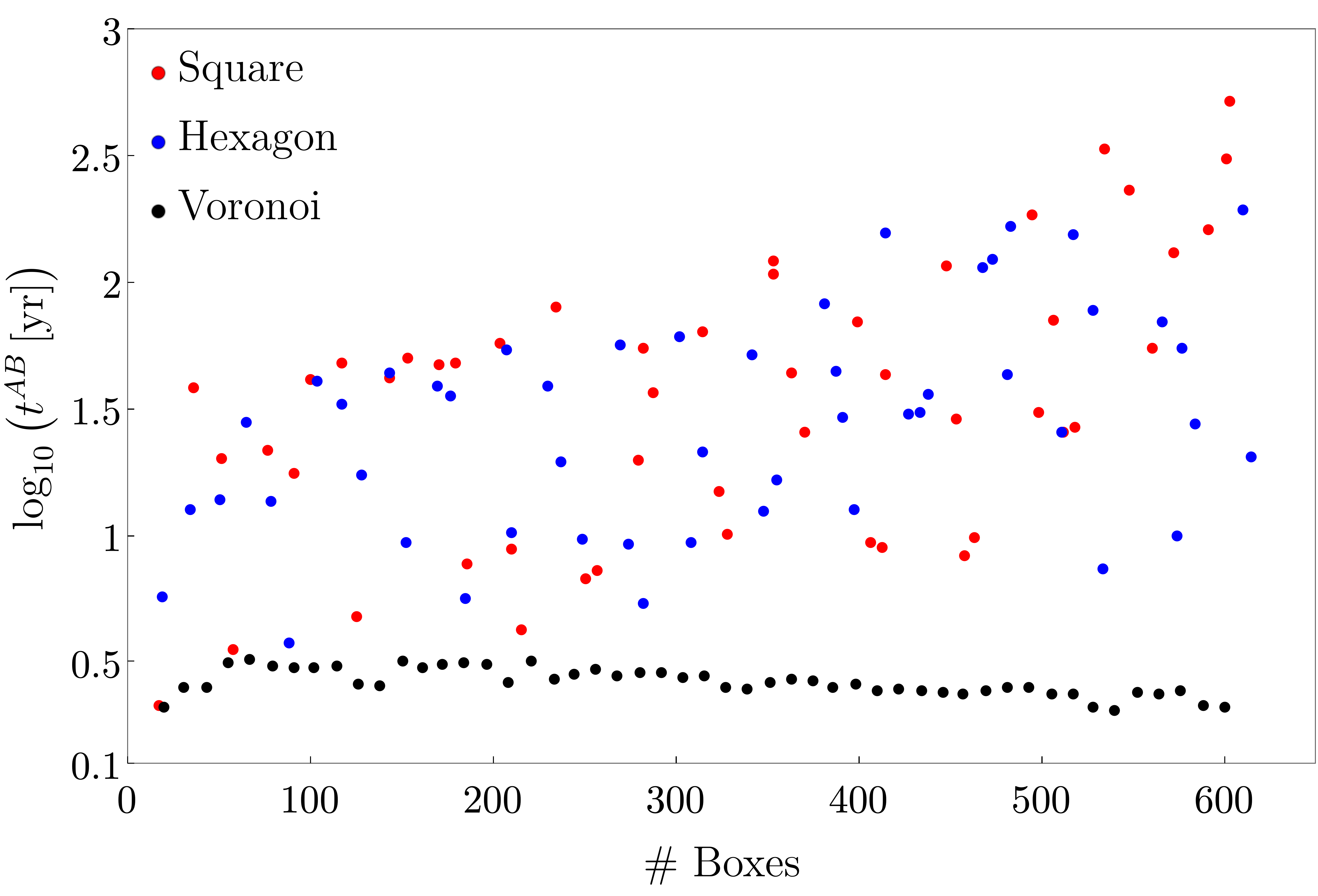}
    \caption{The transition time of Eq.~\eqref{eq:tAB-vanE} for various box sizes and polygon coverings. The horizontal axis shows the number of boxes remaining after boxes with no data were removed, e.g., 600 boxes for a square covering is the result of an initial covering of 1000 boxes.}
    \label{fig:tABvsBOX}
\end{figure}

For this covering, we find $t^{\isset{AB}} = 2.34\,\text{yr}$. Note that Fig.~\ref{fig:mu-vor-500} does not contain any disconnected polygons, and in general the Voronoi covering is less prone to disconnections although they are still possible. To compare the stability of this method to the regular coverings discussed previously, we compute $t^{\isset{AB}}$ for a number of boxes sizes between 20 and 600 as shown in Fig.~\ref{fig:tABvsBOX}.
We see that the Voronoi covering produces significantly more stable transition times of consistently reasonable magnitudes. To understand this, we note that the addition of a small number of clusters does not tend to have a large global effect as observed for regular coverings. Requesting more clusters tends to subdivide larger clusters or create more where the data is dense and leave others untouched. 

The Voronoi covering has some drawbacks. First, although the k-means algorithm is a relatively fast clustering algorithm and amenable to parallelization if necessary, it is still roughly two orders of magnitude slower computationally than the regular coverings. In addition, the initial guess for the location of centroids in the typical k-means algorithm is random. We find that the variation in $t^{\isset{AB}}$ caused by variations in this initial guess are much smaller than those coming from the change in box size. We recommend running the clustering algorithm a handful of times to check that the initial guess has not accidentally found an undesirable local maximum. Finally, we note that the nature of the algorithm means that it is difficult to add boxes in specific locations, but this also applies for non-adaptive regular coverings. 

\subsection{The time step \texorpdfstring{$T$}{T}}
\label{subsec:the-time-step-T}

\begin{figure}[t!]
    \centering
    \includegraphics[width = \linewidth]{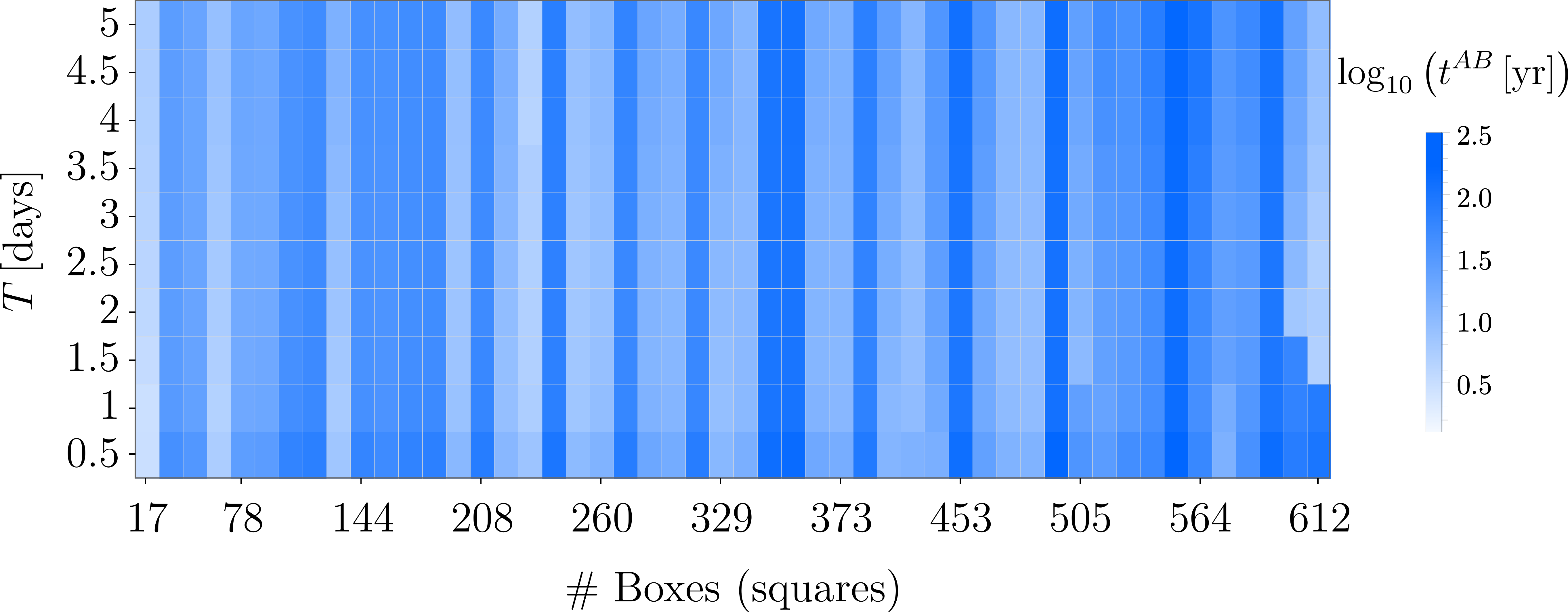}
    \\ \vspace{0.5cm}
    \includegraphics[width = \linewidth]{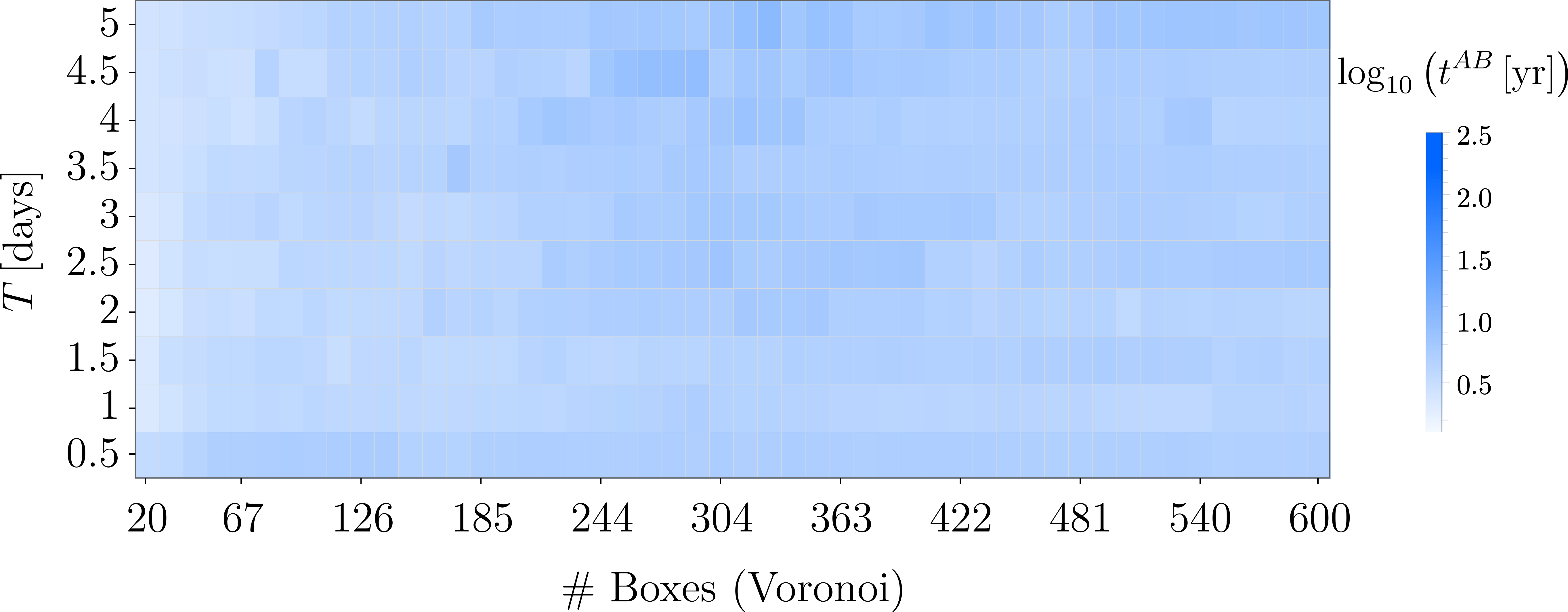}
    \caption{(top panel) The transition time of Eq.~\eqref{eq:tAB-vanE} with a regular covering of squares for various box sizes and time steps. The horizontal axis shows the number of boxes after boxes with no data removed. For a regular hexagonal covering, the graph looks very similar. (bottom panel) As in the top panel, but using the Voronoi covering described in Section~\ref{subsec:voronoi-coverings}. }
    \label{fig:tABboxT-vor}
\end{figure}

As mentioned in Section~\ref{subsec:regular-coverings}, the time step $T = 5\,\text{days}$ was chosen based on time scales arising from the Lagrangian characteristics (decorrelation) in the upper ocean. Here we explore the validity of this choice based on the stability of the transition time. Figure~\ref{fig:tABboxT-vor} shows this transition time as a function of both box number and time step $T$ both for a regular square covering and the Voronoi covering.
Reading across the horizontal axis for a fixed $T$, we see the same results as in Fig.~\ref{fig:tABvsBOX}, namely, that the transition time is stable against box number for the Voronoi covering but not for the regular covering. For a fixed box number, reading up the vertical axis generally shows that $T = 0.5$ days tends to have a slightly higher $t^{\isset{AB}}$ but for $T > 0.5$ days, there is very little variation for both the regular and Voronoi coverings. When $T = 0.5$ days, there are enough trajectories that do not leave their initial cells that the transition matrix is very strongly diagonal; this serves to increase the transition time. As discussed previously, small changes in the number of boxes for a regular covering can result in global shifts in the locations of boxes in the covering. However, small changes in $T$ do not generally have this behaviour since increasing $T$ still leaves the same number of observations (modulo a small number of points left off at the end) and hence for sufficiently long trajectories, the effect is not felt to a significant degree. With real data, there will of course be an upper limit to $T$ beyond which results cease to become trustworthy due to lack of communication between boxes and large numbers of short trajectories being rejected. What we observe here, regardless of considerations related to the Lagrangian decorrelation time of the ocean, is that the lower limit for obtaining stable transition times is roughly 1 day. 

\section{A Generalized transition time} 
\label{sec:generalized-transition-time}

\begin{figure}[t!]
    \centering
    \includegraphics[width = \linewidth]{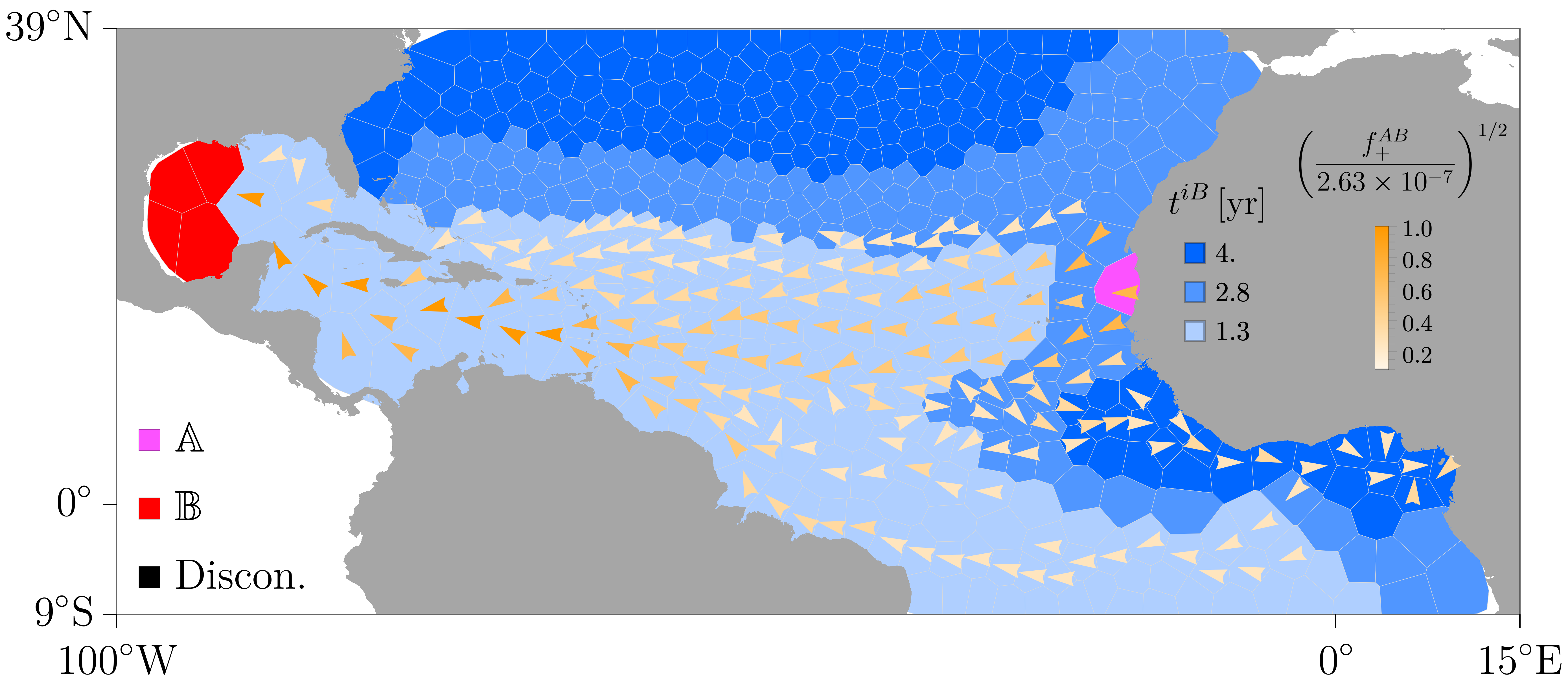}
    \caption{The \trem{} of Eq.~\eqref{eq:trem_def} for a Voronoi covering with 500 cells, partitioned into three regions via k-means clustering, with the effective reactive current of Eq.~\eqref{eq:effective-reactive-current} overlaid.}
    \label{fig:tremfplus}
\end{figure}

In this section, we present a generalization of Eq.~\eqref{eq:tAB-vanE} which can be used to obtain a partition of the computational domain based on the time it takes to reach $\isset{B}$ from an arbitrary cell. The aim of this generalization is to provide local information, that is, information about reactive trajectories at a particular state that have already left the source $\isset{A}$. Let
\begin{equation}
    \isset{C}^+ = \left\{i \notin \isset{B} \st  \sum_{\ell \in \isset{S} }P_{i \ell} q_{\ell}^+ > 0 \right\}.
\end{equation}
We define the \emph{\trem{}}  $t^{i\isset{B}}$ for all $n \in \isset{Z}$ as
\begin{equation} \label{eq:trem_def}
    t^{i\isset{B}} \defn \begin{cases}
    \expec\left[\tau_{\isset{B}}^+(n + 1) \mid X_{n} = i,\ R^{+}(n + 1)\right] & i \notin \isset{B},\\
    0 & i \in \isset{B}.
    \end{cases}
\end{equation}
A similar formula is referred to as the \emph{lead time} in \citet{Finkel-etal-21}. In Appendix~\ref{sec:proofs}, we establish the following Lemma.
\begin{lemma} \label{lemma:tAB}
Eq.~\eqref{eq:trem_def} satisfies a set of linear equations,
\begin{equation} 
    t^{i\isset{B}} \defn \begin{cases}
    1 + \sum_{j \in \isset{C}^+ }\frac{P_{i j} q_{j}^+  }{\sum_{\ell \in \isset{S} }P_{i \ell} q_{\ell}^+} t^{jB} & i \in \isset{C}^+,\\
    0 & i \in \isset{B}.
    \end{cases}
    \label{eq:trem_sys}
\end{equation}
When  $\isset{A}$ contains only one state, we also have that
\begin{equation}
    t^{i\isset{B}} \big|_{i = \isset{A}} = t^{\isset{AB}} + 1,
\end{equation}
where $t^{\isset{AB}}$ is defined in Eq.~\eqref{eq:tAB-vanE}.
\end{lemma}
By applying Lemma~\ref{lemma:tAB}, we can compute the \trem{} for each box in a given covering. We choose a Voronoi covering with 500 boxes, similar to the construction of Fig.~\ref{fig:mu-vor-500}. To build a \emph{remaining-time-based dynamical geography}, we partition the \trem{}s into three clusters via k-means. We show this geography overlaid with the effective reactive current of Eq.~\eqref{eq:effective-reactive-current} in Fig..~\ref{fig:tremfplus}. Due to the large size of the regions, the \trem{}s assigned to them should be taken as representative of these regions and not of any particular cell. This is especially apparent when considering cells near the source or target, e.g., cells in the Gulf of Mexico have \trem{}s on the order of 3-4 weeks. The dynamical geography obtained here is similar to the one obtained in \citet{Beron-etal-22-ADV} by other means.
We see that the longest times are found near the Gulf of Guinea and the most subtropical North Atlantic. This is consistent with the drifter data: there is a large inflow to the Gulf of Guinea, making drifters near the West coast of Africa cause a large pile-up of trajectories in this region. Similarly, the Gulf Stream pushes drifters up and out of the Gulf of Mexico such that they are unlikely to transition back into the Gulf in a short time once they pass the coast of Florida. Consequently, the dynamical geography provided by the \trem{} is related to the distance between states and the target, but they are not interchangeable; the \trem{} provides additional information. 

Examining the reactive current, we recover the two main transition paths observed in \citet{Beron-etal-22-ADV}, namely, the direct westward path and the indirect path which initially moves toward the Gulf of Guinea before circulating westward across the equatorial Atlantic. Not only does the direct path have a larger effective current, the transition times are also shorter. In general, there is a noticeable separation between westward and eastward-bound currents south of the source $\isset{A}$. We note here that the \trem{} of a state need not be positively correlated with its effective current. For example, the effective currents are roughly equal in the Gulf of Guinea and northern portion of the Gulf of Mexico but the \trem{}s are significantly different. 

\section{Conclusions}
\label{sec:conclusions}

When Ulam's method is applied to trajectory data, the space must be partitioned into a covering to discretize the motion and thereby construct a transition probability matrix. We have shown that two types of standard coverings made up of regular grids of squares and hexagons result in unstable transition times when Transition Path Theory (TPT) is applied to the induced Markov chain. Changing the number of squares or hexagons in the covering leads to global shifts in their location, producing untrustworthy results for TPT statistics. We proposed a different kind of covering which partitions the space into Voronoi cells based on k-means clustering of the observations. This covering leads to transition times which are stable against the number of requested clusters. This algorithm was chosen for simplicity and effectiveness, but there are many clustering algorithms which could be explored, in particular with consideration toward improving computational performance for large data sets. In addition, we found that the transition time does not depend strongly on the time step through which the trajectory data are sliced for time steps between 1 and 5 days for undrogued drifters in the tropical/subtropical Atlantic. Finally, we introduced a generalization of the standard TPT transition time, which contains the standard TPT transition time as a special case. Clustering cells based on this generalized transition time produces a partition of the domain which reveals weakly dynamically connected regions.

\appendix

\section{Proof of Lemma~\ref{lemma:tAB}}
\label{sec:proofs}
We first establish that Eq.~\eqref{eq:trem_def} can be written as the solution to the system of linear equations in Eq.\@~\eqref{eq:trem_sys}. In what follows, we repeatedly use the Markov property and the stationarity of our chain. First, we have that
\begin{widetext}
\begin{align}
    \prob(X_{n + 1} = j \mid X_{n} = i,\, R^{+}(n + 1)) &= \frac{\prob(X_{n + 1} = j,\, R^{+}(n + 1) \mid X_{n} = i)}{\prob(R^{+}(n + 1) \mid X_{n} = i)} \\ 
    &= \frac{P_{i j} q_{j}^+ }{\sum_{\ell \in \isset{S}}P_{i \ell} q_{\ell}^+} \label{eq:tAB-P-denom}.
\end{align}
\end{widetext}
Note that the condition that our Markov chain is ergodic does not necessarily imply that $\sum_{\ell \in S }P_{i \ell} q_{\ell}^+ \neq 0$ for all $i \notin \isset{B}$. There can exist a series of states for which the only path between them and $\isset{B}$ passes through $\isset{A}$. Hence, an ``interior'' state whose neighbors all have $q_{\ell}^+ = 0$ will have $\sum_{\ell \in S }P_{i \ell} q_{\ell}^+ \neq 0$. To address this case, we introduce the set of states 
\begin{equation}
    \isset{C}^+ = \left\{i \notin \isset{B} \st  \sum_{\ell \in \isset{S}}P_{i \ell} q_{\ell}^+ > 0 \right\}
\end{equation}
and restrict $t^{i\isset{B}}$ to these states. Therefore, Eq.~\eqref{eq:tAB-P-denom} is well-defined. Taking $i \notin \isset{B}$ in Eq.~\eqref{eq:trem_def}, we condition on the value of $X_{n + 1}$ and use the fact that $\tau_{B}^+(n + 1) = 1 + \tau_B^+(n + 2)$ on the event that $X_{n + 1} \notin \isset{B}$ to obtain 
\begin{equation} 
    t^{i\isset{B}} \defn \begin{cases}
    1 + \sum_{j \in \isset{C}^+ }\frac{P_{i j} q_{j}^+  }{\sum_{\ell \in \isset{S}}P_{i \ell} q_{\ell}^+} t^{jB} & i \in \isset{C}^+ \\
    0 & i \in \isset{B}
    \end{cases} .
\end{equation}
We will now show that Eq.~\eqref{eq:trem_def} with $i = \isset{A}$ coincides with Eq.~\eqref{eq:tAB-vanE}. We compute the quantity $Z(n) = \prob(R^{-}(n), R^{+}(n))$. For the time $n$ to be reactive, there must be a last visit to $\isset{A}$ at some time $\ell < n$, and the next visit to $\isset{A}\cup \isset{B}$ must be to $\isset{B}$ at least $n - \ell$ steps later. Therefore, we can write
\begin{widetext}
\begin{align}
    Z(n) &= \sum_{\ell < n} \prob\left(X_\ell \in \isset{A}, R^{+}(\ell + 1), \tau_{\isset{B}}^+(\ell + 1) > n - \ell\right) \\ 
    &= \sum_{\ell < n} \prob\left(\tau_{\isset{B}}^+(\ell + 1) \geq n - \ell \mid X_\ell \in \isset{A},\, R^{+}(\ell + 1)\right) \prob\left(X_\ell \in \isset{A}, R^{+}(\ell + 1)\right)
\end{align}
\end{widetext}
By the stationarity of our process, $\prob(X_\ell \in \isset{A}, R^{+}(\ell + 1))$ is independent of $\ell$, so we have
\begin{widetext}
\begin{equation}
    Z(n) = \prob\left(X_n \in \isset{A},\, R^{+}(n + 1)\right) \sum_{\ell < n} \prob\left(\tau_{\isset{B}}^+(\ell + 1) > n - \ell \mid X_\ell \in \isset{A},\, R^{+}(\ell + 1)\right).
\end{equation}
\end{widetext}
Applying the stationary property again and re-indexing the sum over $\ell$ gives
\begin{widetext}
\begin{equation}
      Z(n) = \prob\left(X_n \in \isset{A},\, R^{+}(n + 1)\right) \sum_{\ell  = 1}^{\infty} \prob( \tau_{\isset{B}}^+(n + 1) > \ell \mid X_n \in \isset{A}, R^{+}(n + 1)).
\end{equation}
\end{widetext}
Since for a nonnegative discrete random variable $X$ we have $\expec[X] = \sum_{k \geq 1} P(X \geq k)$, we conclude that
\begin{widetext}
\begin{equation}
    Z(n) = \prob\left(X_n \in \isset{A},\, R^{+}(n + 1)\right)\left(-1 + \expec\left[\tau_{\isset{B}}^+(n + 1) \mid X_{n} = i, R^{+}(n + 1)\right]\right),
\end{equation}
\end{widetext}
which implies that Vanden-Eijden\citep{VandenEijnden-06}'s $t^{\isset{AB}}$ and Eq.~\eqref{eq:trem_def} with $i = \isset{A}$ differ by one step, i.e., they are identical except that $t^{\isset{AB}}$ does not ``count'' the first step to leave $\isset{A}$.

\begin{acknowledgments}
The authors are grateful to Luzie Helfmann for providing notes which inspired the development of Eq.~\eqref{eq:trem_def}. This work was supported by the National Science Foundation under grant OCE2148499.
\end{acknowledgments}

\section*{Author declarations}

\subsection*{Conflict of Interest}

The authors have no conflicts to disclose.

\subsection*{Author Contributions}

\textbf{Gage Bonner:} Conceptualization (equal); Formal analysis (lead); Computation (lead); Visualization (lead); Software (lead); Writing – original draft (lead); Writing – review \& editing (equal). \textbf{Francisco Javier Beron-Vera:} Conceptualization (equal); Writing – review \& editing (equal); Funding acquisition (equal). \textbf{Maria Josefina Olascoaga:} Conceptualization (equal); Writing – review \& editing (equal); Funding acquisition (equal).

\section*{Data availability}

The data employed in this paper are openly available from the NOAA Global Drifter Program at \href{http://www.aoml.noaa.gov/phod/dac/}{\texttt{http://www.aoml.noaa.gov/phod/dac/}}. The computations were carried out using \texttt{Julia}; a package has been developed, which is distributed from \href{https://github.com/70Gage70/UlamMethod.jl}{\texttt{https://github.com/70Gage70/UlamMethod.jl}}. 

\%bibliography{mybib.bib,fot.bib}
%

\end{document}